\documentclass[%
 reprint,
 amsmath,amssymb,
 aps,
 prl,
twocolumn,
]{revtex4-2}

\usepackage[utf8]{inputenc}
\usepackage{graphicx}
\usepackage{float}
\usepackage{xcolor}
\usepackage{amsmath}
\usepackage{multirow}
\usepackage{array}
\newcolumntype{P}[1]{>{\centering\arraybackslash}p{#1}}
\usepackage{braket}
\usepackage{amsmath,bm}
\usepackage{wrapfig}
\usepackage{hyperref}
\hypersetup{
    colorlinks,
    citecolor=black,
    filecolor=black,
    linkcolor=black,
    urlcolor=black
}

\usepackage{braket}
\usepackage{amsmath,bm}

\begin{document}
\preprint{APS/123-QED}

\title{Building ideal paraxial optical skyrmions using rational maps}

\author{C. Cisowski\textsuperscript{1},  S. Franke-Arnold\textsuperscript{1}, C. Ross\textsuperscript{2,3}}
\affiliation{\textsuperscript{1}School of Physics and Astronomy, University of Glasgow, Glasgow. UK.}
\affiliation{\textsuperscript{2}Department of Mathematics, University College London, London, UK}
\affiliation{\textsuperscript{3}Department of Physics and Research and Education Center for Natural Sciences, Keio
University, Japan}

\date{\today}

\begin{abstract}
We introduce a simple mathematical expression based on rational maps to construct ideal paraxial optical skyrmions fields including Néel-type and Bloch-type  skyrmions, anti-skyrmions, bimerons and multi-skyrmions, including skyrmion lattices. We review the rules that fully polarized paraxial light fields must obey to be considered as optical skyrmions. This work provides guidelines for the experimental generation of general skyrmion fields, beyond conventional single skyrmion beams. This lays the foundation for the exploration of nucleation and annihilation mechanisms in multi-skyrmions fields.
\end{abstract}

\maketitle
\textit{Introduction.}---Skyrmions were introduced as field configurations of minimal energy in a non-linear scalar field model of the atomic nucleus by Tony Skyrme in 1962~\cite{SKYRME1962556}. Skyrme's model is founded on an elegant topological construction~\cite{Manton1987} which confers a universal dimension to skyrmions. Skyrmions, as topological objects, have been identified in various areas of physics including condensed matter physics \cite{han2017}, string theory~\cite{Sakai2005} and magnetism~\cite{Bogdanov2020}. 

Research on optical skyrmions is still in its infancy. First witnessed in the evanescent field of surface plasmon polariton in 2018~\cite{Tsesses2018,Du2019}, optical skyrmions have since been generated in freely propagating light fields in the paraxial~\cite{Gao2020,Shen2022,Sugic2021} and the non-paraxial~\cite{Cuevas_2021} regimes.

The interest for optical skyrmions lies in their topological nature, which could confer considerable stability to these fields if they share this property with their magnetic cousins~\cite{Sampaio2013}. They could also be used to write and read information in ferromagnetic materials~\cite{PhysRevLett.99.047601,Lambert}, with magnetic skyrmions being envisioned to form the basis of the next generation of data storage devices~\cite{Fert2013-fe,FosterDennis2019}. 

Advances in optical skyrmions research have mostly been driven by drawing analogies with their magnetic counterparts, with Néel-type and Bloch-type skyrmions~\cite{Gao2020, Cuevas_2021}, Bimerons and anti-skyrmions~\cite{Shen2022} being introduced as examples of optical skyrmions. These analogies are justified as both paraxial optical skyrmions and magnetic skyrmions are `baby' skyrmions: skyrmion fields defined in a two dimensional space ~\cite{PIETTE1994294,Piette1995}. In the following we will refer to 2D skyrmions as skyrmions for simplicity. 

Presently, a unifying framework capable of describing all optical skyrmions, including multi-skyrmions and skyrmion lattices, is lacking. This letter introduces an analytical expression to construct general paraxial optical skyrmion fields using rational maps. Rational maps have been used to solve Skyrme's non-integrable field equations since 1998~\cite{HOUGHTON1998507}, but have never been employed to construct optical skyrmions. This work provides guidelines for generating novel experimental skyrmion fields and lays the foundations for exploring how bosonic skyrmion fields differ from their magnetic counterparts. Indeed, expressing optical skyrmions in terms of rational maps effectively lifts the veil on the equations optical skyrmions are the solutions of, hereby raising questions on their physical interpretation.

To provide a unified description of optical skyrmions, we examine how skyrmion fields are constructed from topological considerations. This outlines criteria for distinguishing optical skyrmions from other fully polarized paraxial light fields, including Poincar\'e beams.      

\begin{figure*}
    \centering
    \includegraphics[width=0.8\textwidth]{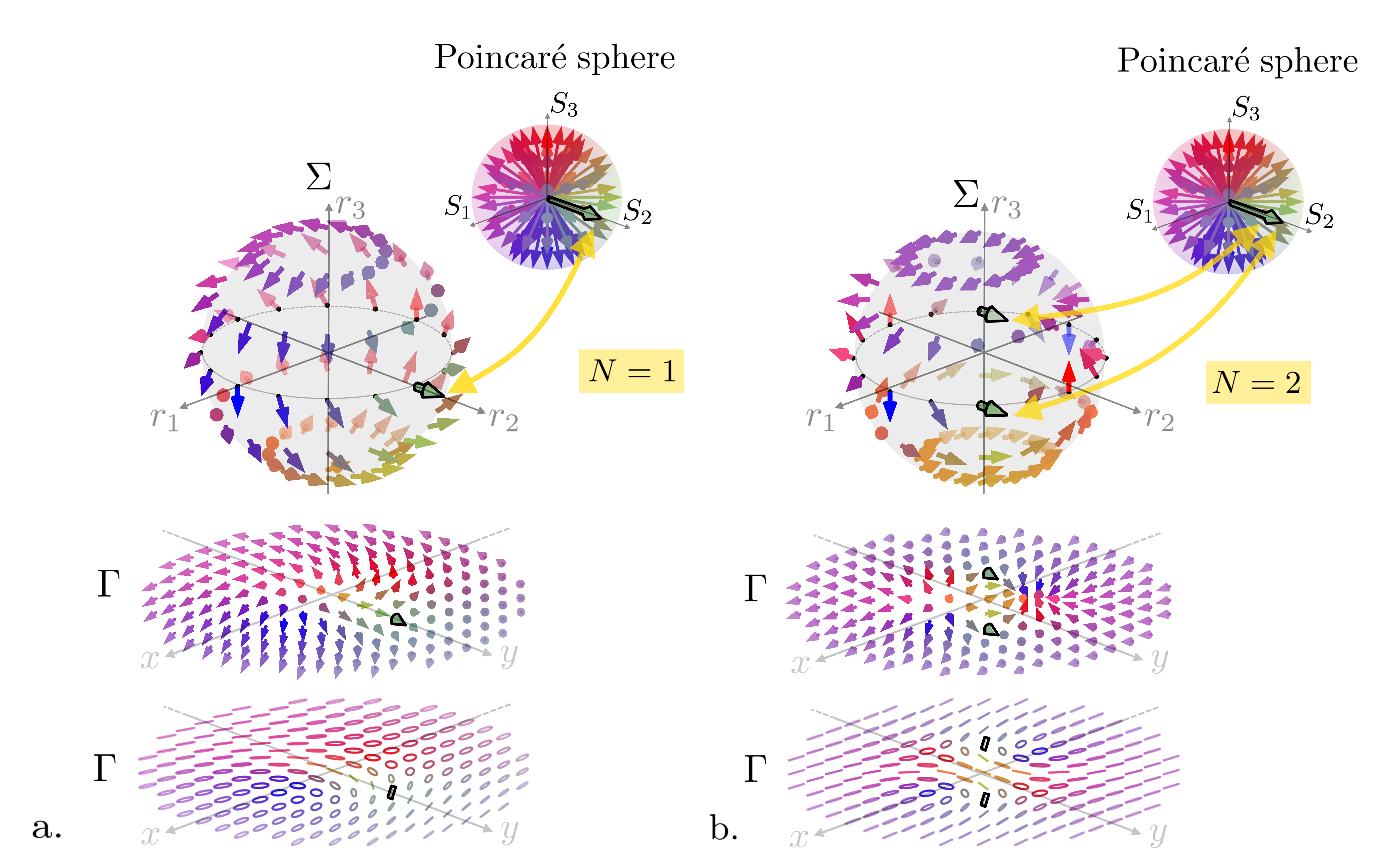}
    \caption{Paraxial optical fields of skyrmion number a. $N=1$ and b. $N=2$. The plane $\Gamma$ has been transformed into a sphere $\Sigma$. For $N=1$ the vector field $\bm{S}(r_1,r_2,r_3)$ on $\Sigma$ covers the Poincar\'ee sphere once whereas for $N=2$ the Poincar\'e sphere is covered twice. Consequently, the $S_2$ vector, highlighted by a bold outline, appears once on $\Sigma$ (hence once in $\Gamma$) for $N=1$ and twice on $\Sigma$ (hence twice in $\Gamma$) for $N=2$. The polarization distribution in $\Gamma$ is given as both a $\bm{S}$ distribution and a distribution of polarization ellipses.}
    \label{fig1}
\end{figure*}

\textit{Topology of 2D skyrmions}---A paraxial optical skyrmion field assigns to each position $(x,y)$ in the plane transverse to the propagation direction of the light beam, we shall call this plane $\Gamma$, a three-component unit vector, the reduced Stokes vector $\bm{S}(x,y)=[S_1(x,y),S_2(x,y),S_3(x,y)]^T$~\cite{Gao2020}. The tip of the vector $\bm{S}(x,y)$ defines a point on the Poincar\'e sphere representing the local polarization state. Depicting the polarization distribution in $\Gamma$ as a vector field $\bm{S}(x,y)$ or as a spatial distribution of polarization ellipses is equivalent. A paraxial optical skyrmion field is thus a map from a plane to a sphere, this is the general definition of 2D skyrmion fields including magnetic skyrmions~\cite{Piette1995}. 
An interesting situation occurs when the plane $\Gamma$ can be transformed into a sphere, which we denote $\Sigma$, in a process called compactification. As illustrated in Fig.~\ref{fig1}, the skyrmion field becomes a map between two spheres and an integer, called skyrmion number $N$ can be associated with this mapping.   
Building a sphere from a plane is no easy task, as cartographers can attest, but it can be accomplished using an inverse stereographic projection. This techniques identifies spatial infinity with a single point on the sphere of unit radius $\Sigma$, then uses this point as a projection point to image all remaining spatial positions $(x,y)$ in the plane $\Gamma$ into points on the sphere $\Sigma$. If, for instance, the projection point is the North Pole of $\Sigma$, $\bm{r}=(r_1,r_2,r_3)=(0,0,1)$, a point $(x,y)$ in $\Gamma$, defined by $r_3=0$, becomes the point $\bm{r}=(\frac{2x}{1+x^2+y^2},\frac{2y}{1+x^2+y^2},\frac{-1+x^2+y^2}{1+x^2+y^2})$
on $\Sigma$. The vector $\bm{S}(x,y)$, attached to a point $(x,y)$ in $\Gamma$ becomes attached to the image of the point, $\bm{r}$, in $\Sigma$.  
For paraxial optical skyrmions, this construction requires that, in the plane $\Gamma$, $\bm{S}(x,y)$, hence the polarization state, is the same in all directions at spatial infinity. This situation occurs naturally in Skyrme's model as it ensures that the energy gradient is divergentless hence that the field energy is finite \cite{manton_sutcliffe_2004}. 
The overall distribution of $\bm{S}(x,y)$ is determined by the mapping between $\Sigma$ and the Poincar\'e sphere, which varies from skyrmion to skyrmion. We shall provide the rules for these mappings shortly but let us first define the 
skyrmion number.      
The skyrmion number $N$ counts the number of times the tips of the vectors $\bm{S}(\bm{r})$ on $\Sigma$ cover the entire Poincar\'e sphere. Fig.~\ref{fig1}.a. shows that for $N=1$, $\bm{S}(\bm{r})$ covers the Poincar\'e sphere once hence the $\bm{S}(x,y)$ distribution in the plane $\Gamma$ contains all possible polarization states exactly once. Fig.~\ref{fig1}.b. shows that for $N=2$, $\bm{S}(\bm{r})$ covers the Poincar\'e sphere twice hence all possible polarization states appear twice in $\Gamma$. 
The skyrmion number can be calculated from $\bm{S}(x,y)$ as:   
\begin{equation}
N=\frac{1}{4\pi}\int_A \bm{S}\cdot\left(\partial_{x}\bm{S}\times\partial_{y}\bm{S}\right)dxdy.
\label{skyrmionnumber}
\end{equation}
Eq.\ref{skyrmionnumber} measures the area on the Poincar\'e sphere covered by $\bm{S}$ as we explore an area $A$ in $\Gamma$ then divides this area by the total area of the Poincar\'e sphere. In practice, the measured skyrmion number is often smaller than the true skyrmion number. This happens when the mapping from $\Gamma$ to $\Sigma$ is incomplete. For ideal skyrmions this happens because we cannot integrate the skyrmion number over an infinite grid. For experimental skyrmions this usually happens because of the presence of noise at the periphery of the beam, which corresponds to a low intensity region. For the simple types of skyrmions, such as N\'eel-type and Bloch-type skyrmions, this issue can easily be identified as the polarization state at the periphery never becomes orthogonal to the polarization state at the centre of the beam. Another factor that can lower the measured skyrmion number is the capacity to capture the variations of the $\bm{S}(x,y)$ field, which becomes more challenging as $N$ increases. This problem can be addressed by either increasing the spatial resolution of the computational grid for ideal skyrmions or by increasing the resolution of the imaging system for experimental skyrmions. Because all paraxial optical skyrmions contain all the polarization states within their transverse profile they are, de facto, Poincar\'e beams \cite{Beckley:10}. The converse, however, is not true. Firstly, not all Poincar\'e beams see their polarization state tend to the same state towards spatial infinity. Secondly, skyrmions are constructed according to mapping rules, which may not be necessarily the case for Poincar\'e beams. \\
%
\begin{figure*}
    \centering
    \includegraphics[width=0.9\textwidth]{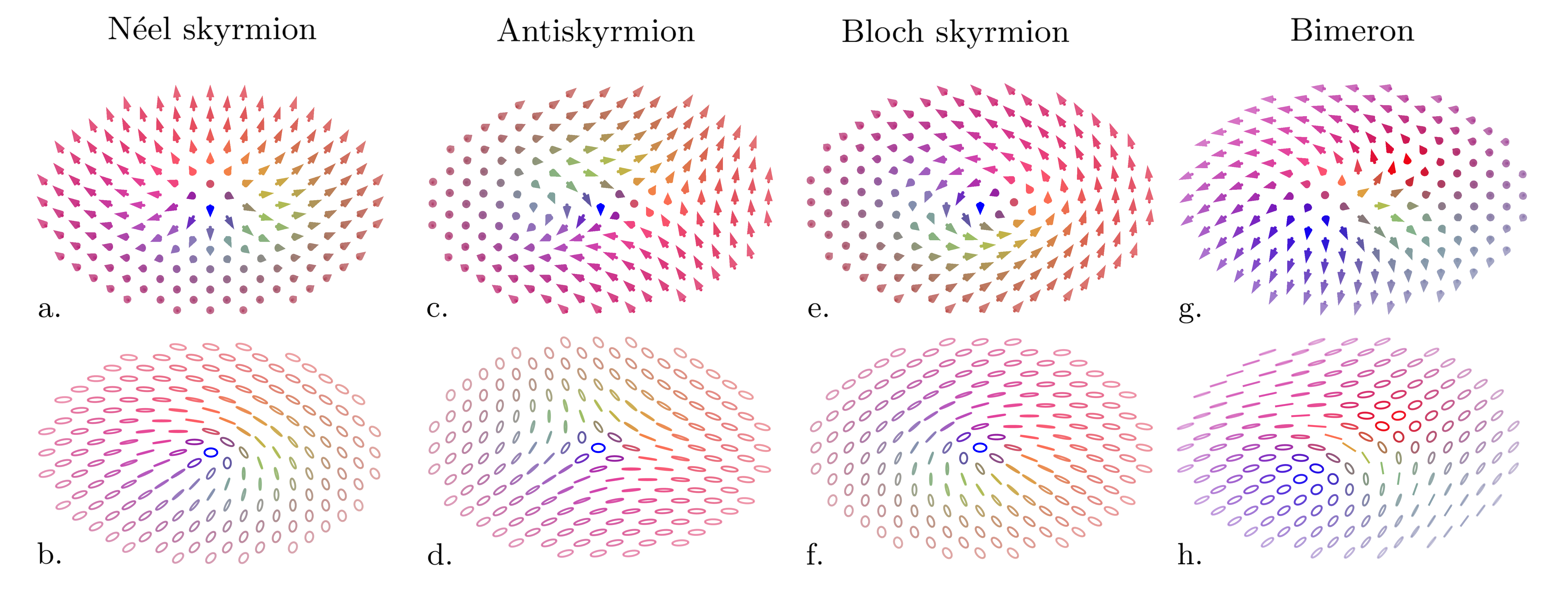}
    \caption{Polarization distribution as a vector field $\bm{S}(x,y)$ and as distribution of polarization ellipses for a N\'eel-type skyrmion (a,b), an antiskyrmion (c,d), a Bloch-type skyrmion (e,f) and a bimeron (g,h) obtained using Eq.\ref{Req}.}
    \label{fig2}
\end{figure*}
\textit{Single skyrmions}---We shall now provide the rules for the mapping between $\Sigma$ and the Poincar\'e sphere to construct the simplest type of optical paraxial skyrmions,  which includes N\'eel-type and Bloch-type skyrmions, anti-skyrmions and bimerons.
Our approach is based on rational maps.   
Rational maps entered the world of skyrmions in 1998 \cite{HOUGHTON1998507}. In Skyrme's original model, a skyrmion is a field solution of the equation that minimizes the energy functional of the system. Skyrme's field equation is not integrable therefore one must often resort to numerical techniques to find its solutions. However, computing the minimum energy configurations using the energy functional can be time-consuming and require significant computational resources. To facilitate this task, Houghton et al. proposed to use rational map to find approximate solutions  \cite{HOUGHTON1998507}. This method is in fact rather accurate, especially for skyrmions of low skyrmion number \cite{dBATTYE}. Rational maps were used to construct Bogomolny-Prasad-Sommerfield (BPS) monopoles prior to skyrmions \cite{Donaldson}. It is by noting that BPS monopoles and low energy skyrmions possess similar energy densities, symmetry and spatial distributions that Houghton et al. proposed to use rational maps to build skyrmions. While Skyrme's field equation describe skyrmions in three dimensions, rational maps are
equally capable of describing 2D skyrmion fields~\cite{Polyakov:1975yp,Hen2008}.
For this reason, we can use rational maps to construct optical skyrmions. A rational map is a function linking two Riemann spheres $R:S_1^2\rightarrow S_2^2$. Considering $z$, the stereographic coordinate on $S_1^{2}$, we define a rational map of degree $D$ as:  
\begin{equation}\label{Req}
    R(z)=\frac{p(z)}{q(z)},
\end{equation}
where $D=\rm{max}(deg(p),deg(q))$. For paraxial optical skyrmions we wish to establish a correspondence between $\Sigma$ and the Poincar\'e sphere. We identify $z$ with the complex coordinate $z=x+iy$ in the plane $\Gamma$ via a stereographic projection through the North pole of $\Sigma$, the unit vector $\bm{S}$ assigned to each $(x,y)$ becomes:
\begin{equation}\label{ansatz}
\bm{S}=\frac{1}{1+|R(z)|^2}
(2\,\textrm{Re}(R(z)),2\,\textrm{Im}(R(z)),-1+|R(z)|^2), 
\end{equation}
N\'eel-type skyrmions, of skyrmion number $N$, can then be built using rational maps of the form: 
\begin{equation}\label{eqrat}
R(z)=z^N, 
\end{equation}
The $\bm{S}(x,y)$ distribution obtained using Eq.~\ref{eqrat} is shown in Fig.~\ref{fig2}.a,b. Using the complex conjugate of $z=x+iy$ in Eq.~\ref{eqrat} yields a polarization distributions that correspond to antiskyrmions (see Fig.~\ref{fig2}.c.,d.), which are skyrmions of negative skyrmion number \cite{Kovalev2018}. This is because complex conjugation swaps the orientation of the skyrmion field. Bloch-type skyrmions, such as the one shown in Fig.~\ref{fig2}.e.,f., or any intermediate skyrmion for which the polarization state is right handed circularly polarized at infinity and left handed circularly polarized at the origin can be obtained by rotating $z$ using $\textrm{exp}(i\alpha)z$ where $\alpha=\pi/2$ for a Bloch-type skyrmion. A rotation of the vector $\bm{S}$ generates skyrmion fields presenting arbitrary orthogonal polarization states at infinity and at the center of the plane. 
In particular, an anti-clockwise $\pi/2$ rotation around the $S_2$-axis will generate an in-plane skyrmion, also called a bimeron~\cite{Shen2022}. 
As illustrated in Fig.\ref{fig2}.g.,h., bimerons present orthogonal linearly polarized states at infinity and at the center. 
From a geometric point of view, a rotation of $\bm{S} $ simply induces a rotation of the Poincar\'e sphere hence does not change how many times $\bm{S}$ covers the Poincar\'e sphere. The freedom to independently rotate $\bm{S}$ and $z=x+iy$ is a feature of the standard Skyrme model. 
In the context of the nuclear Skyrme model these are referred to as rotations, for the domain (here $\Sigma$), and isorotations, for the target (here Poincar\'e sphere) \cite{Manton2022}. Interestingly, the ability to perform these independent rotations is lost in magnetic skyrmions due to a term in the total energy of the system called Dzyaloshinskii-Moriya energy, this term being invariant only under a combination of rotations and isorotations. It does not seem to be the case for optical skyrmions. The details of the Dzyaloshinskii-Moriya energy depend on the specifics of the underlying model which restricts the allowed configurations. In magnetism, it stems from a combined effect of spin–orbit coupling and broken inversion symmetry \cite{Treves1962}. The explicit rational maps, defined in Eq.~\ref{eqrat}, that can describe paraxial optical skyrmions, correspond to the minimum energy configurations in the $O(3)$-sigma model. They solve the equations $\partial_{z}\varphi(x,y)=0$ or $\partial_{\bar{z}}\varphi(x,y)=0$, where $\varphi(x,y)=\frac{S_1+iS_2}{1-S_3}$ and where $\bar{\varphi}$ indicates complex conjugation, with the Skyrme number positive for the first case and negative for the second. Uncovering the equations which paraxial optical skyrmions are the solutions of is of paramount importance as it constitutes the first step in understanding how optical skyrmion fields differ from their magnetic counterparts. \\ 
Fig.~\ref{fig:Skyrmion3} compares the polarization distribution of N\'eel-type skyrmions obtained using Eq.~\ref{eqrat}, with the polarization distribution obtained by superposing two orthogonally polarized Laguerre Gaussian (LG) beams, which is currently the method used to generate paraxial optical skyrmion beams experimentally \cite{Gao2020}.
We consider the following beam superposition:
\begin{equation}
    \Psi = \frac{1}{\sqrt{2}}\left( \rm{LG}_{0}^{0}\ket{0}+\rm{LG}_{0}^{\ell_1}\ket{1}\right),
    \label{Eq:beamsup}
\end{equation}
where $\ket{i}$ with $i \in \{0,1\}$ represent left handed circularly polarized light and right handed circularly polarized light, respectively and where $\rm{LG}_p^{\ell}$ are Laguerre Gaussian modes of radial order $p$ and topological charge $\ell$. There is good agreement between both polarization distributions, which confirms the validity of our model. Skyrmions built using rational maps are ideal skyrmions, in the sense that they are deduced directly from the mapping linking the Poincar\'e sphere to $\Sigma$ (hence $\Gamma$), Eq.\ref{Eq:beamsup}, however, takes into consideration the experimental realization of these beams. 
A few minor differences can thus be noted, firstly, Eq.~\ref{Eq:beamsup} aims to perform the full $\Sigma$ mapping within a confined region of space, secondly, the radial dependence of $\rm{LG}_p^{\ell}$ beams affects the polarization distribution at the centre of the beam. In Fig~\ref{fig:Skyrmion3}.h., for instance, the region of left-handed elliptically polarized polarization (in blue) is larger than the one of Fig~\ref{fig:Skyrmion3}.g., this discrepancy increases with $N$.        
\begin{figure}
    \centering
    \includegraphics[width=0.5\textwidth]{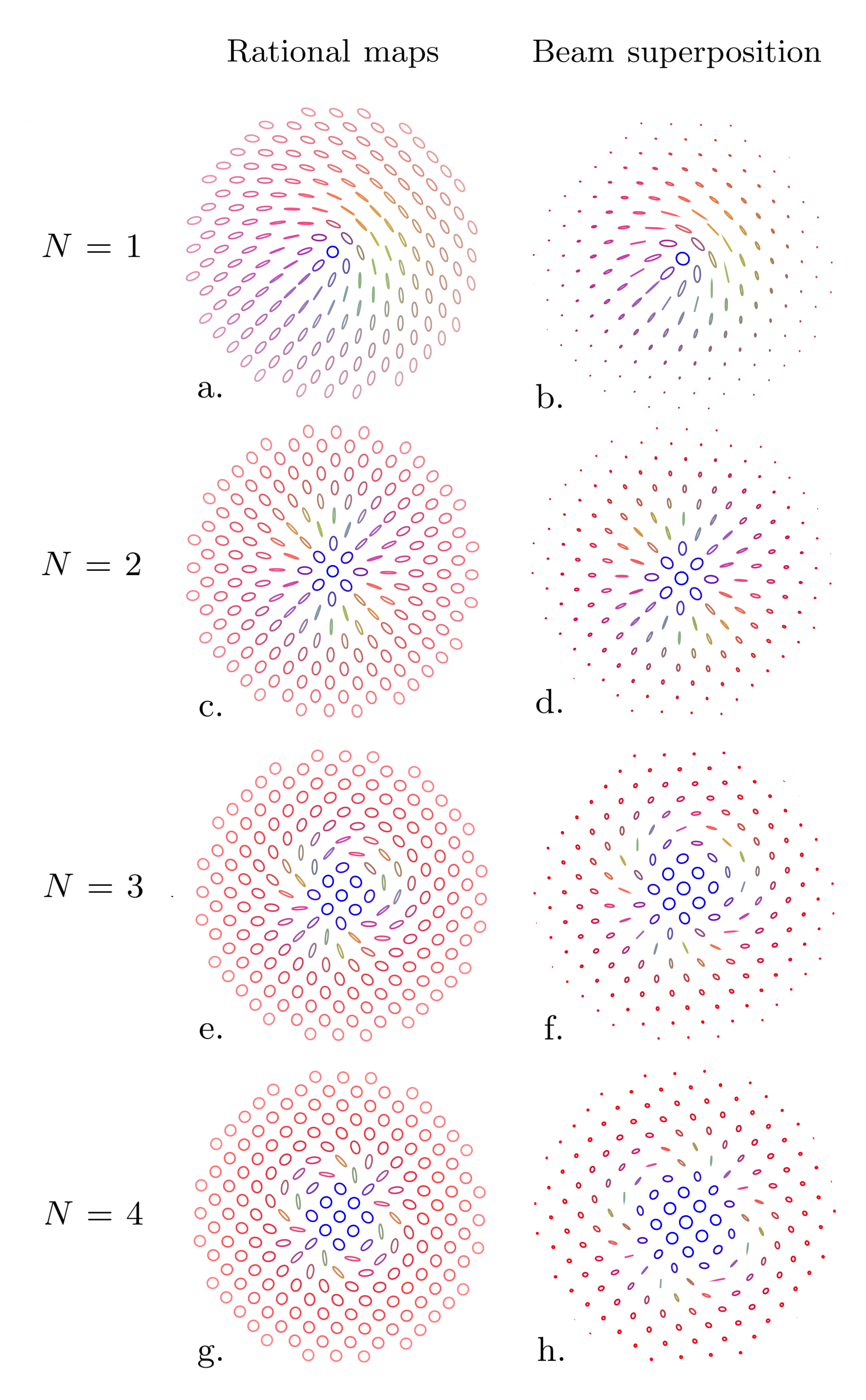}
    \caption{Polarization distributions, illustrated in terms of polarization ellipses, of paraxial optical skyrmions of skyrmion number $N=1$ (a,b), $N=2$ (c,d), $N=3$ (e,f), $N=4$ (g,h). The skyrmions shown in (a,c,e,g) were obtained using Eq.\ref{eqrat} whereas the field shown in (b,d,f,h) were obtained with Eq.~\ref{Eq:beamsup} for $\ell_1=1,2,3,4$, respectively. In the second case, the ellipse size is modulated by the intensity profile of the beam.} 
    \label{fig:Skyrmion3}
\end{figure}
All optical skyrmions constructed using Eq.~\ref{eqrat} are ``single skyrmions": $R(z)$ present one zero where $z=0$. Eq.~\ref{eqrat} introduces a unifying framework for N\'eel type and Block type skyrmions, antiskyrmions and bimerons. Beam superposition methods to construct different types of optical single skyrmions have been outlined in \cite{Shen2022}. The construction of optical multi-skyrmions, however, is not straightforward, unless one knows the rules for the $\Gamma$ to $\Sigma$ mapping. We shall now see how Eq.~\ref{eqrat} can be generalized to design multi-skyrmion beams including skyrmion lattices. 

\begin{figure}[h!]
    \centering
    \includegraphics[width=0.5\textwidth]{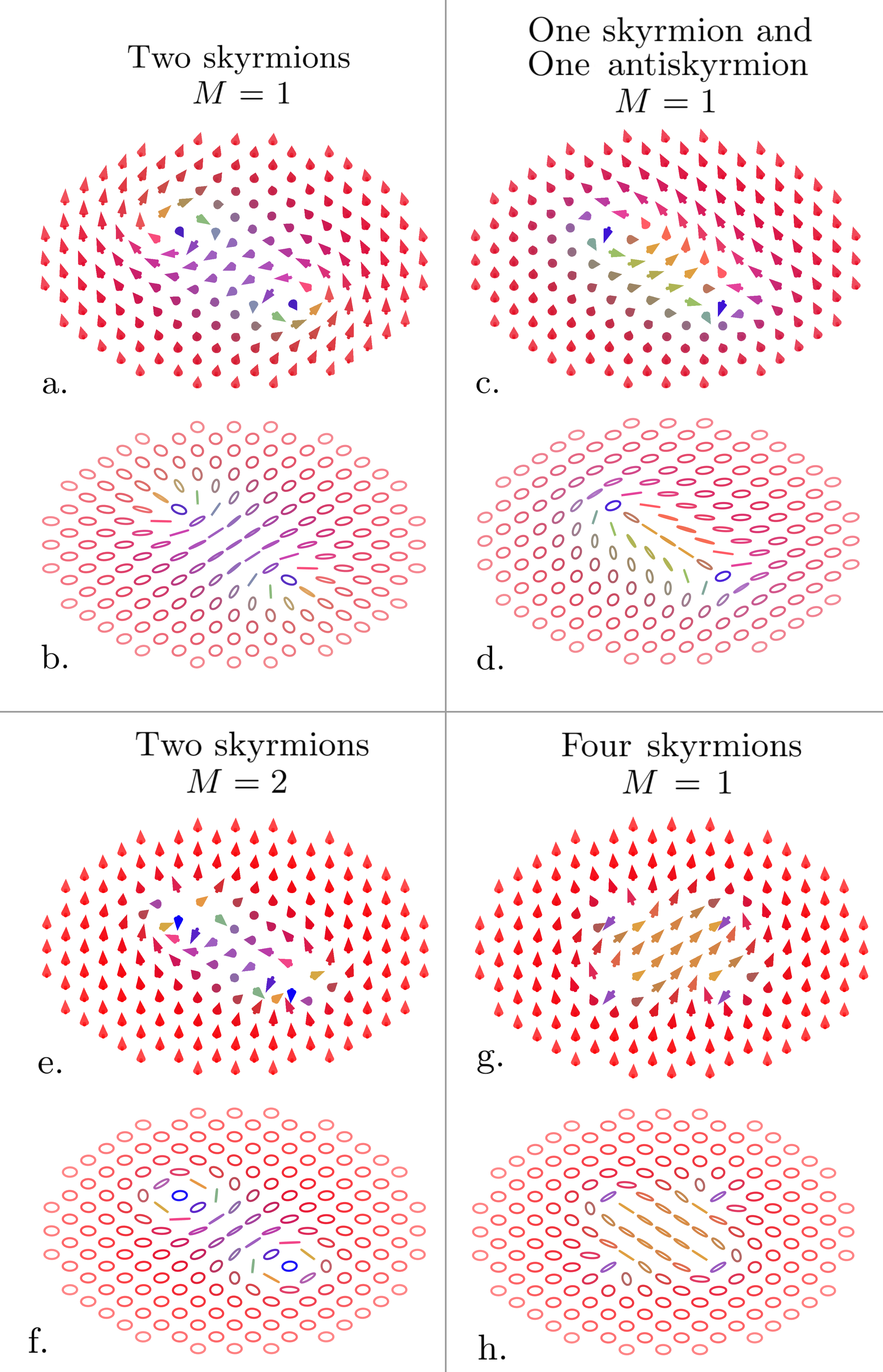}
    \caption{Polarization distributions, as vector fields (a,c,e,g) and polarization ellipses (b,d,f,h) of two skyrmions zeros $M=1$ (a,b), one skyrmion zero and one antiskyrmion zero $M=1$ (c,d), two skyrmions zeros with $M=2$ (e,f) and four skyrmions zeros whith $M=1$ (g,h).}
    \label{fig4}
\end{figure}

\textit{Multi-skyrmions}--Ideal, paraxial multi-skyrmion beams can be constructed by generalizing Eq.~\ref{eqrat} such that:
\begin{equation}
R(z)= \prod (x-dx_j\pm i(y-dy_j))^{M},
\label{Eq:}    
\end{equation}
where $dx_j$ ad $dy_j$ are the x and y position in $\Gamma$ of the $j^{\rm{th}}$ zero and where the $\pm$ distinguishes skyrmions from antiskyrmions, respectively. $M$ is an integer that corresponds to the absolute value of the skyrmion number of the individual zero. Fig.~\ref{fig4} shows the polarization distribution of two skyrmion zeros where $dx_{1,2}=0, dy_{1}=1, dy_{2}=-1$ and $M=1$ (Fig.~\ref{fig4}.a.,b.), one skyrmion zero and one anti skyrmion  where $dx_{1,2}=0, dy_{1}=1, dy_{2}=-1$ and $M=1$ (Fig.~\ref{fig4}.c.,d.), two skyrmion zeros where  $dx_{1,2}=0, dy_{1}=1, dy_{2}=-1$ and $M=2$ (Fig.~\ref{fig4}.e.,f.) and four skyrmion zeros where $dx_{1,2}=-1, dx_{3,4}=+1, dy_{1,3}=-1, dy_{2,4}=+1$ and $M=1$ (Fig.~\ref{fig4}.g,h.). The polarization distribution obtained in figure~\ref{fig4}.a,b. is similar to the one of a magnetic biskyrmion such as the one presented in \cite{Yu2014,Gbel2019} to describe two partially overlapping magnetic skyrmions. By providing an analytical expression for multi-skyrmions, we introduce guidelines for the experimental generation of multi-skyrmion fields, including skyrmion lattices. Annihilation and creation mechanisms of optical skyrmion zeros would also constitute a new line of research. Indeed, each zero contributes to the the total skyrmion number by an amount given by the degree of the rational map $M$. This suggests that we can split an optical skyrmion of skyrmion number $N$ into several skyrmions zeros of total skyrmion number $N=\sum M$ or obtain a skyrmion number of $N=0$ by carefully combining skyrmions zeros with antiskyrmions zeros. These effects may be witnessed by perturbing vortex beams, these investigations will likely be based on works concerned with polarization singularity dynamics \cite{Otte2018, DErrico2017}.      

\textit{Concluding remarks}--- We have shown that rational maps can be used to build ideal paraxial optical skyrmion fields. Not only can they model single optical skyrmion fields, including N\'eel type and Bloch type skyrmions, antiskyrmions and bimerons, but they can also be used to construct muti-skyrmions, paving the way for the experimental realization of general optical skyrmions fields and enabling the study of skyrmion-skyrmion interactions in optical fields. By interpreting optical skyrmions fields in terms of rational maps, we find that optical skyrmions corresponds to minimum energy configurations of the $O(3)$-sigma model and that, unlike magnetic skyrmions, they appear to be invariant to rotations and isorotations. This invites for further investigations regarding the physical significance of the $O(3)$-sigma model for bosons. This will entail confirming or ruling out the existence of stabilizing terms for optical skyrmion fields, which should be possible by studying their behaviour upon perturbations and free propagation.  

\begin{acknowledgments}
C.M.C. acknowledges financial support from the Royal Society through an International Newton fellowship NIF/R1/192384.
\end{acknowledgments}

\bibliography{apssamp}

\end{document}